\documentclass[runningheads]{llncs}
\usepackage[T1]{fontenc}
\usepackage{graphicx}
\usepackage{subcaption}
\usepackage{enumitem}
\usepackage{caption}
\usepackage{orcidlink}
\usepackage{todonotes}
\usepackage{wrapfig}
\usepackage{xspace}
\usepackage{booktabs}
\usepackage{siunitx}
\begin{document}
\title{An Explainable AI Assistant for Introductory Programming Education: Improving Feedback Reliability with Instructor-AI Collaboration}
%
\titlerunning{An Explainable AI Assistant for Introductory Programming Education}
%

\author{
Muntasir Hoq\inst{1}\orcidlink{0000-0003-2591-0476} \and
Griffin Pitts\inst{1}\orcidlink{0009-0004-3111-6118} \and
Bradford Mott\inst{1}\orcidlink{0000-0003-3303-4699} \and
Seung Lee\inst{1}\orcidlink{0000-0002-7782-2331} \and \\
Jessica Vandenberg\inst{1}\orcidlink{0000-0001-6497-1840} \and
Shuyin Jiao\inst{1}\orcidlink{0000-0002-0621-1605} \and
Narges Norouzi\inst{2}\orcidlink{0000-0001-9861-7540} \and \\
James Lester\inst{1}\orcidlink{0000-0003-1481-6601} \and 
Bita Akram\inst{1}\orcidlink{0000-0001-5195-5841}
}

\authorrunning{M. Hoq et al.}

\institute{
North Carolina State University, Raleigh, NC, USA \\
\email{\{mhoq, wgpitts, bwmott, sylee, jvanden2, sjiao2, lester, bakram\}@ncsu.edu}
\and
University of California, Berkeley, CA, USA \\
\email{norouzi@berkeley.edu}
}
\maketitle              
\begin{abstract}
Active learning is widely recognized as an effective approach for improving learning outcomes in introductory programming courses. However, insufficient instructional support often limits students' access to timely, personalized feedback, which is crucial for mastering foundational programming concepts. Although recent advances in AI, particularly large language models, offer scalable opportunities for feedback, concerns about explainability and reliability remain. In this paper, we present an AI-driven classroom assistant that leverages an explainable AI model to analyze student code, map logical errors to instructor-identified misconceptions, and deliver instructor-authored feedback, thereby grounding reliability in instructor-defined pedagogical knowledge. To evaluate the effectiveness of our framework, we conducted an expert evaluation to examine its alignment with instructor-verified feedback and deployed the system in a classroom setting to assess students' perceptions of its usability. Results indicate that the assistant can provide accurate, instructor-verified feedback to students while fostering a positive experience.

\end{abstract}
\section{Introduction} 
In introductory programming courses, active learning emphasizes in-class activities where students engage with and apply course material \cite{mcconnell1996active}. Such activities can include problem-solving, discussion, and writing code to support student engagement and conceptual understanding during class. During active learning, timely and individualized feedback helps students identify and correct misconceptions as they learn to program \cite{denny2024desirable}. However, the diversity of students' solutions and the large number of students in introductory programming courses make it difficult for instructors to provide such feedback at scale during class time, limiting the effectiveness of active learning \cite{tang2024sphere}.

Among emerging approaches for addressing this challenge, large language models (LLMs) offer a scalable mechanism for analyzing student programs and producing adaptive feedback \cite{phung2023generating,jia2024llm}. Despite their potential, LLMs raise concerns about reliability and trustworthiness, particularly in introductory settings where misconceptions can persist if introduced early \cite{chen2020impact}. Further, formative interviews with instructors in our prior work suggested hesitation about exposing novice students to unverified LLM-generated feedback, as well as concerns about whether such feedback would align with the goals of active learning~\cite{hoq2025facilitating}.

To address these challenges, we propose a triangulated approach that integrates instructor expertise, LLMs, and an explainable AI model (SANN~\cite{hoq2025automated}) to provide reliable, instructor-verified feedback at scale. This approach is instantiated in the \textsc{Insight} classroom assistant, which supports instructor--AI collaboration through an authoring tool that enables instructors to define common student misconceptions and associated feedback. The framework matches student code to these instructor-authored patterns and delivers targeted feedback aligned with focal instructional goals.
Through \textsc{Insight}, student feedback takes the form of short, instructor-authored natural language messages that explain an error and suggest revisions. LLMs are not used to generate feedback shown directly to students. Instead, they play constrained roles in assisting instructors during authoring, generating synthetic code for model fine-tuning, and verifying when student errors fall outside instructor-defined coverage.

We evaluate our approach using four programming problems from the FalconCode dataset~\cite{de2023falconcode}. \textsc{Insight} was assessed on real student submissions through a human evaluation comparing model-selected feedback to expert instructor judgments and LLM-generated feedback. Results demonstrate that instructor-AI collaboration can provide reliable, pedagogically grounded feedback aligned with instructional goals. In addition, an initial classroom deployment indicated positive student perceptions of the system in an authentic instructional setting.

This paper makes the following contributions:

\begin{itemize}
    
    \item Proposing a novel feedback propagation framework that combines an explainable code analysis model, instructor-verified error–feedback pairs, and constrained LLM support to select reliable feedback for students.
    \item Developing a training pipeline that adapts to newly authored problems using only synthetic code, mirroring classroom settings where prior student data is not available.
    \item Evaluating the framework on multiple FalconCode problems and, through an initial classroom deployment, demonstrating high-quality feedback matching for instructor-prioritized misconceptions and positive student perceptions of clarity and usefulness.


\end{itemize}

\section{Related Work}

Introductory programming courses often use active learning activities that engage students in problem-solving, discussion, and writing code during class~\cite{chi2014icap,mcconnell1996active}. These approaches can improve understanding of core concepts, but are most effective when students also receive timely, individualized feedback to diagnose and correct misconceptions~\cite{denny2024desirable}. Because instructor attention during class time is limited, providing this level of feedback at scale remains a major challenge~\cite{tang2024sphere}.

Automated feedback systems can address this challenge by analyzing student code and providing targeted feedback to help students correct errors~\cite{messer2024automated}. Early systems focused on test-based grading or automated repair~\cite{singh2013automated,gulwani2018automated,bhatia2018neuro,gupta2019deep,zhang2022repairing}, offering limited natural language explanations and often emphasizing correctness over pedagogical value. Moreover, static code analysis methods are often used for prediction tasks or recommending similar worked examples, but not for providing specific feedback to students on current incorrect submissions~\cite{hoq2023sann,hoq2025worked}. Recent work explores leveraging LLMs to generate feedback on syntax errors, bugs, and misconceptions~\cite{phung2023generating,jia2024llm,phung2024automating,jacobs2024evaluating,koutcheme2024open}. While LLMs can produce rich natural language, they raise concerns around hallucinations, lack of explainability, and misalignment with instructional intent~\cite{phung2023generative,jacobs2024evaluating,jia2024assessing}. Our work responds to these challenges by combining an explainable program representation model with instructor-authored feedback and tightly constrained LLM support to provide scalable, trustworthy feedback aligned with focal instructional points.

\section{\textsc{INSIGHT} Classroom Assistant} \label{classroom assistant}

    

To support scalable, reliable, and explainable feedback delivery, we developed \textsc{Insight}, a web-based classroom assistant that facilitates active learning following an iterative co-design process with instructors \cite{hoq2026insight}. \textsc{Insight} supports programming exercise authoring, distribution, and adaptive support. This paper focuses on its AI-driven framework, which selects and propagates adaptive feedback based on student-created code. The system includes an instructor-facing interface and a student-facing interface. 

\begin{figure}[t]
    \centering
    \fbox{\includegraphics[width=0.65\columnwidth]{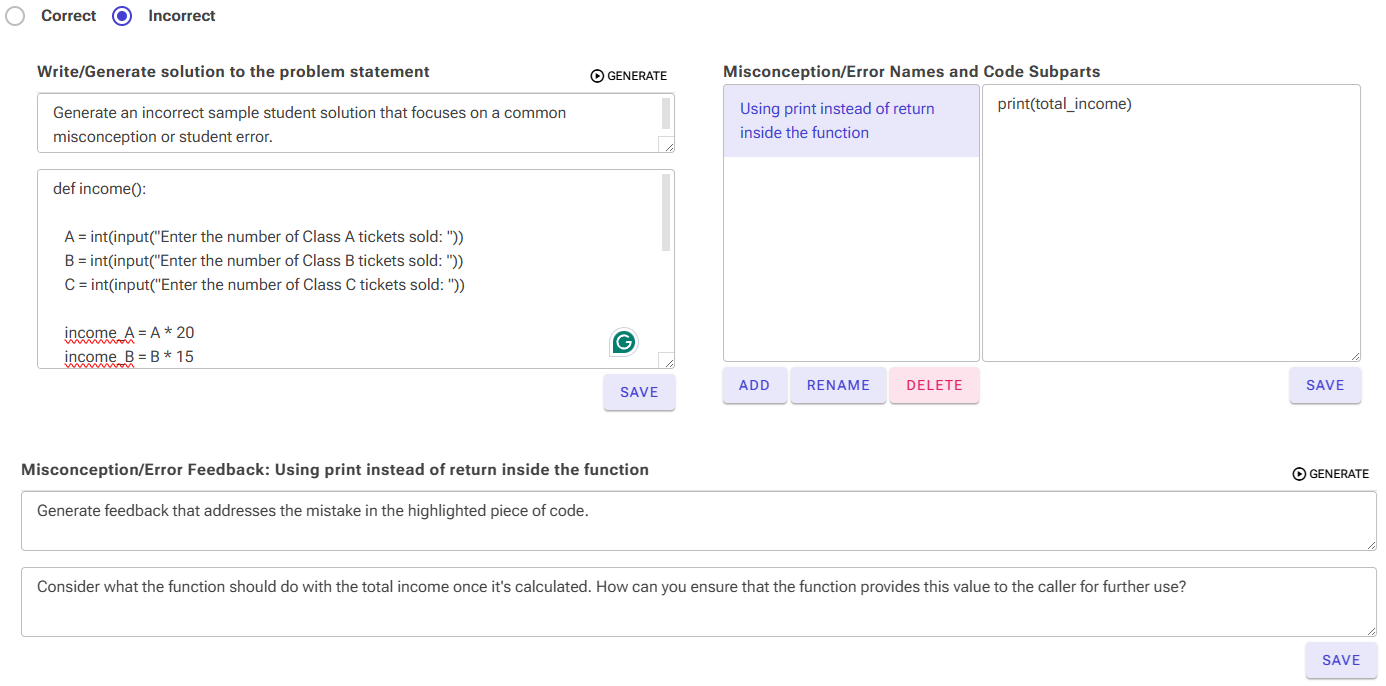}}
    \caption{Misconception and feedback tagging interface in the authoring tool.}
    \label{fig:instructor_tool3}
\end{figure}

The instructor interface includes a dashboard for assigning exercises and an authoring tool for creating exercises, specifying correct and incorrect solutions, and attaching feedback to instructionally relevant misconceptions (Figure~\ref{fig:instructor_tool3}). Instructors can optionally collaborate with an LLM to generate new exercises, common student misconceptions, and corresponding feedback. 

\begin{figure}[t]
    \centering
    \includegraphics[width=0.67\columnwidth]{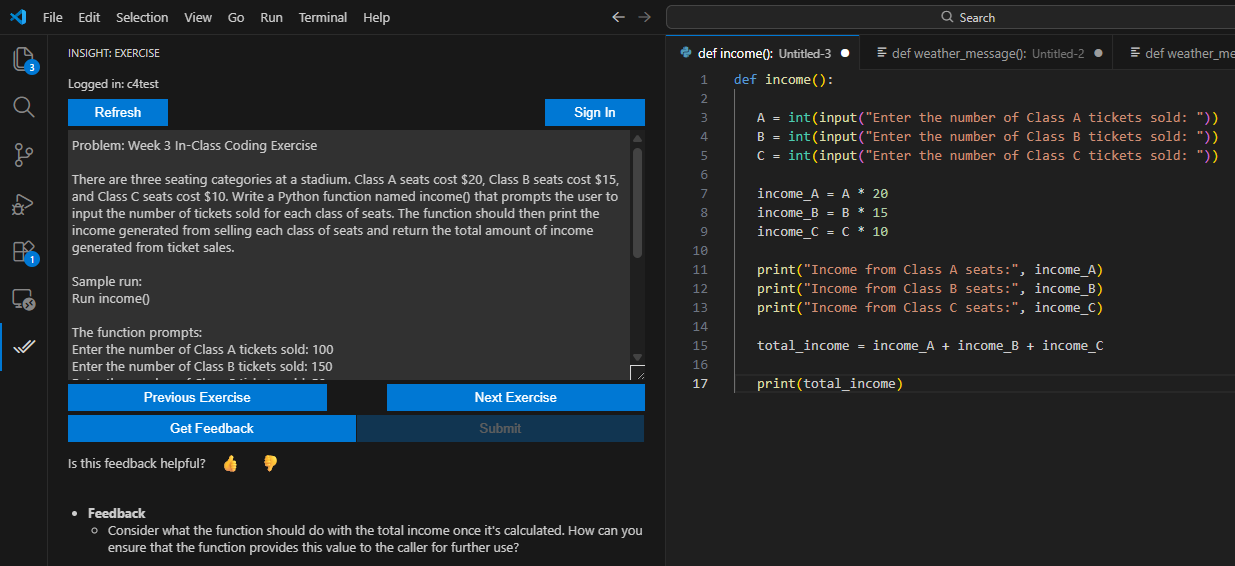}
    \caption{The \textsc{Insight} VS Code extension as the student interface.}
    \label{fig:studentapp}
\end{figure}
The student interface (Figure~\ref{fig:studentapp}), implemented as a Visual Studio Code extension, enables students to receive exercises, submit solutions, and view feedback directly in their development environment, while logging all interactions to a cloud backend.

\section{Feedback Framework and Offline Evaluation}

\begin{figure}[ht]
    \centering
    \includegraphics[width=0.9\textwidth]{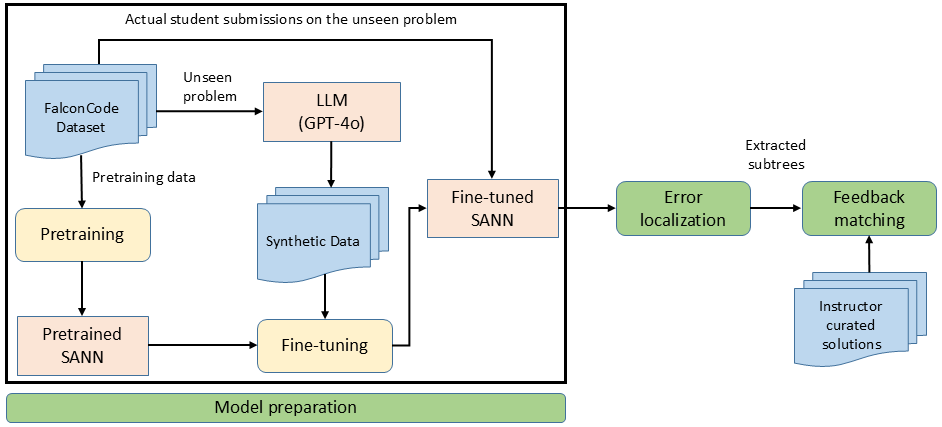}
    \caption{Feedback propagation framework.}
    \label{fig:feedback_prop}
\end{figure}

Our feedback propagation framework combines an explainable AI model with LLM-based support to deliver accurate and reliable feedback on student programs. As shown in Figure~\ref{fig:feedback_prop}, the framework consists of three stages: (1) model preparation, where our explainable AI model (SANN) is pretrained on historical submissions and then fine-tuned on synthetic data for a newly authored problem; (2) error localization, where the fine-tuned SANN identifies student submission logical errors by identifying influential Abstract Syntax Tree (AST) subtrees using an attentional mechanism\cite{hoq2025automated}; and (3) feedback matching, where error subtrees are matched against instructor-authored examples and the designated feedback is returned to the student.

\subsection{Dataset}
For this work, we used the FalconCode dataset~\cite{de2023falconcode}, a large, publicly available collection of $1.5$ million Python programs submitted by over $2,000$ undergraduate students at the US Air Force Academy over $5$ semesters. The dataset includes code submissions for more than 800 Python programming assignments, along with metadata such as problem prompts, test cases used for evaluation, and the specific skills required to solve each problem. The problems cover fundamental CS topics, such as conditionals, loops, files, functions, strings, and lists.



\subsection{Model Preparation}
\label{sec:model_training}
\subsubsection{Subtree-based Attention Neural Network (SANN).}
Subtree-based Attention Neural Network (SANN)~\cite{hoq2025automated} is an explainable program representation learning model that encodes source code into compact vectors capturing syntactic and semantic information. It extracts subtrees from the AST, embeds them into vectors, and aggregates them into a single program representation using an attention network, assigning scalar weights to highlight the most influential structures. These representations have been effectively applied to various prediction tasks, e.g., code correctness prediction and algorithm detection~\cite{hoq2023sann,hoq2024detecting}. In \textsc{Insight}, we leverage SANN’s attention mechanism to identify key subtrees that contribute to the model’s prediction of program correctness, enabling us to focus feedback on the specific regions of incorrect code.

\subsubsection{Model Training.}
To ensure the model can identify a variety of errors in student code, we pretrained the SANN model using the FalconCode dataset on a binary correctness classification task. 
To streamline the pretraining process and make it less resource-hungry, we randomly selected $50$\% of the available problems, resulting in a dataset of $234$ problems and over $557,377$ student submissions ($113,038$ correct and $444,339$ incorrect). Consistent with prior research~\cite{hoq2025automated}, $121,763$ incorrect, uncompilable solutions were excluded from the dataset as they cannot be parsed into ASTs. Given the inherent class imbalance, a stratified $80$:$10$:$10$ split was performed for train, validation, and test sets to preserve the label distribution across all subsets. Hyperparameters were tuned based on validation performance. We set the embedding dimension to $100$, the number of AST nodes per subtree, and the number of extracted subtrees per code to $100$, and applied an early stopping patience of $20$ epochs within a maximum of $200$ training epochs. On the held-out test set, the pretrained model achieved an accuracy of $88$\%, precision of $88$\%, recall of $81$\%, and an F1-score of $83$\%, indicating that it captures syntactic and semantic patterns relevant to program correctness and is suitable as a foundation for fine-tuning and feedback selection.


\begin{figure}[h]
    \centering
    \begin{subfigure}[b]{0.4\textwidth}
        \includegraphics[width=\linewidth]{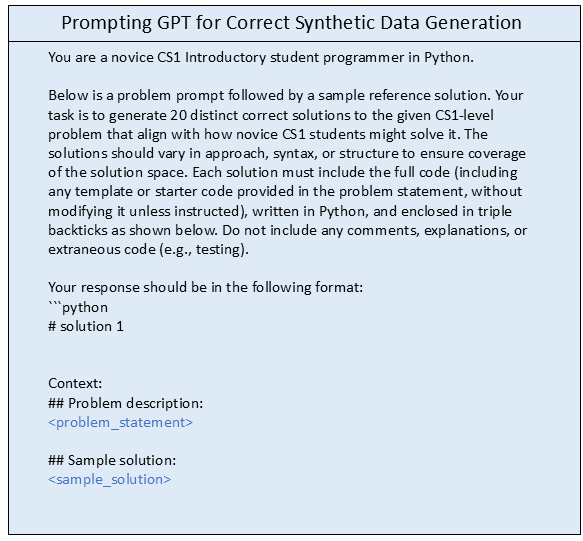}
        \label{fig:correct}
    \end{subfigure}
    \begin{subfigure}[b]{0.4\textwidth}
        \includegraphics[width=\linewidth]{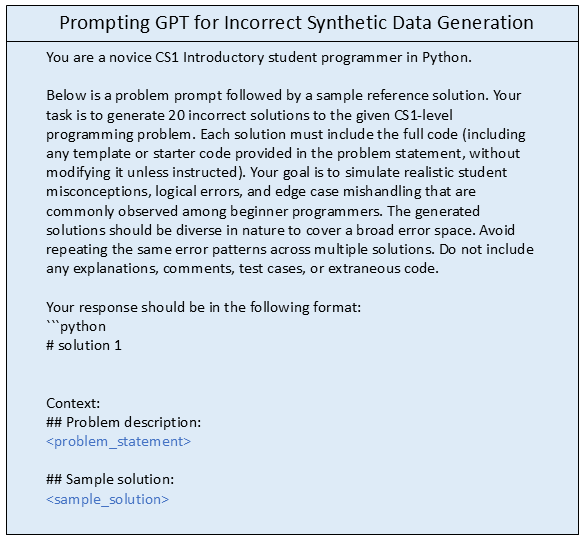}
        \label{fig:incorrect}
    \end{subfigure}

    \caption{Prompt for correct and incorrect synthetic code generation using GPT.}
    \label{fig:synthetic_data}
\end{figure}

To evaluate the model's ability to generalize to unseen problems, we randomly selected a problem focused on conditionals from the FalconCode dataset that was not included in pretraining, mirroring classroom scenarios where new problems lack historical student data. Using the \textsc{Insight} authoring interface, an instructor authored two correct solutions and a set of incorrect solutions representing common misconceptions for this problem. We then leveraged GPT-4o to generate synthetic submissions: $1{,}000$ correct and $1{,}000$ incorrect programs, using a structured prompt instructing the model to mimic introductory student programmers. The prompt emphasized diversity in correct implementations and common logical errors, conceptual misunderstandings, and edge-case mishandling in incorrect ones, and included the instructor-generated sample solutions.
To increase variety, we adopted a multi-sampling strategy: for each prompt, we requested 20 distinct solutions and repeated this process 50 times for both correct and incorrect sets 
(Figure~\ref{fig:synthetic_data}). This yielded a diverse collection of synthetic code that represents plausible student responses to the problem. The pretrained model was fine-tuned on this synthetic data to adapt to the new problem without any historical student submissions.

To assess the effectiveness of our fine-tuned model, we evaluated its performance at predicting the correctness of the $548$ real student submissions in the dataset (correct: $168$, incorrect: $380$) for the same unseen problem used in the fine-tuning phase. We compared our approach (fine-tuned SANN) against three baselines: (i) a SANN model trained only on the synthetic data without pretraining (synthetic-only SANN), (ii) the pretrained SANN model applied directly to the new problem without fine-tuning, and 
(iii) GPT-4o. As shown in Table~\ref{tab:correctness}, the fine-tuned SANN outperformed all the baselines significantly ($p$-$value$ < $0.05$ using Wilcoxon signed rank test) across accuracy, precision, recall, and F1-score, demonstrating the advantage of using a pretrained model adapted specifically for unseen problems through LLM-generated synthetic data. 

\begin{table}[t]
\centering
\caption{Program correctness prediction performance on an unseen problem.}
\label{tab:correctness}
\begin{tabular}{l
                S[table-format=1.2]
                S[table-format=1.2]
                S[table-format=1.2]
                S[table-format=1.2]}
\toprule
\textbf{Model} &
\textbf{Accuracy} &
\textbf{Precision} &
\textbf{Recall} &
\textbf{F1-score} \\
\midrule
Synthetic-only SANN & 0.83 & 0.81 & 0.76 & 0.78 \\
Pretrained SANN     & 0.70 & 0.35 & 0.50 & 0.41 \\
GPT-4o              & 0.86 & 0.85 & 0.83 & 0.83 \\
\midrule
\textbf{Fine-tuned SANN} & \bfseries 0.89 & \bfseries 0.89 & \bfseries 0.84 & \bfseries 0.86 \\
\bottomrule
\end{tabular}
\end{table}


\subsection{Error Localization}
We adopt the error-localization approach of~\cite{hoq2025automated} to identify likely logical errors in student submissions. After a student submits code, a SANN model fine-tuned on synthetic data for the target problem analyzes the submission and assigns attention scores to AST subtrees. We select subtrees whose attention exceeds a $15$\% threshold and treat them as the regions most associated with the predicted incorrectness. This threshold was chosen empirically as a conservative cutoff to retain only highly attended subtrees while avoiding excessive inclusion of low-attention regions. These localized subtrees provide a basis for selecting feedback that targets the relevant parts of an incorrect solution. 


\subsection{Feedback Selection}

For each problem, the instructor authors a set of incorrect example solutions, annotates the relevant erroneous code regions, labels the underlying misconception, and provides associated feedback. We apply the same subtree localization procedure to these instructor examples to build a library of feedback-linked error patterns. At runtime, the fine-tuned model localizes influential subtrees in each student submission, and we compute cosine similarity between the vector representations of student error regions and those from the instructor library.

Feedback is then assigned to the incorrect student submissions based on the closest matching erroneous subtrees from the instructor examples. If the highest similarity score exceeded a threshold of $50$\%, the feedback associated with the nearest instructor-defined error pattern is delivered to the student; otherwise, no feedback is provided, ensuring precise and contextually relevant guidance. To further reduce false positives for errors not represented in the instructor examples, we include an additional GPT-4o verification layer~\cite{koutcheme2024open}.

\subsubsection{Feedback Selection Performance.}

We evaluated feedback-matching quality on the fine-tuned problem through a human evaluation conducted by two expert evaluators. The evaluation focused on matching accuracy, whether the feedback corresponded appropriately to the actual student error (precision), rather than pedagogical quality, since all feedback was instructor-authored. The evaluators also checked whether any feedback on code errors was missing (recall). First, both evaluators independently labeled 10\% of the student submissions, achieving high inter-rater agreement (Cohen’s $\kappa = 0.89$~\cite{landis1977measurement}). After resolving disagreements, one evaluator completed the remaining annotations. Prior to evaluation, submissions containing syntax or runtime errors were excluded, resulting in $169$ incorrect programs, as these errors are generally identified during execution and receive automatic feedback from the programming environment, allowing our evaluation to focus on more nuanced logical and conceptual issues prioritized by the instructor. 


Quantitative analysis showed that among all errors from incorrect submissions, $63$\% received correctly matched feedback, $23$\% received incorrect feedback, and $14$\% received no feedback due to low similarity or unmatched patterns. To gain additional insight,  we conducted further analysis that revealed that the majority of submissions in the ``incorrect'' and ``no feedback'' categories deviated substantially from the problem specification or fell outside the scope of instructor-authored examples. For instance, the required boilerplate code for user input was removed from the submission. 

To limit the propagation of incorrect feedback, we applied the GPT-4o verification layer described earlier~\cite{koutcheme2024open}. The model was asked to assess, as a judge, whether the matched feedback is appropriate for the identified student code region and consistent with the instructor-defined error set. When the verification step judged the feedback to be irrelevant, the submission was reclassified into the ``no feedback'' category. In practice, such cases would be flagged for instructor follow-up rather than receiving automated guidance. This additional filtering reduced the rate of incorrect feedback from $23$\% to $1.2$\%, thereby enhancing the overall reliability of the feedback propagation framework. 

\begin{wrapfigure}{r}{0.55\textwidth}
    \vspace{-\baselineskip} 
    \centering
    \includegraphics[width=0.55\textwidth]{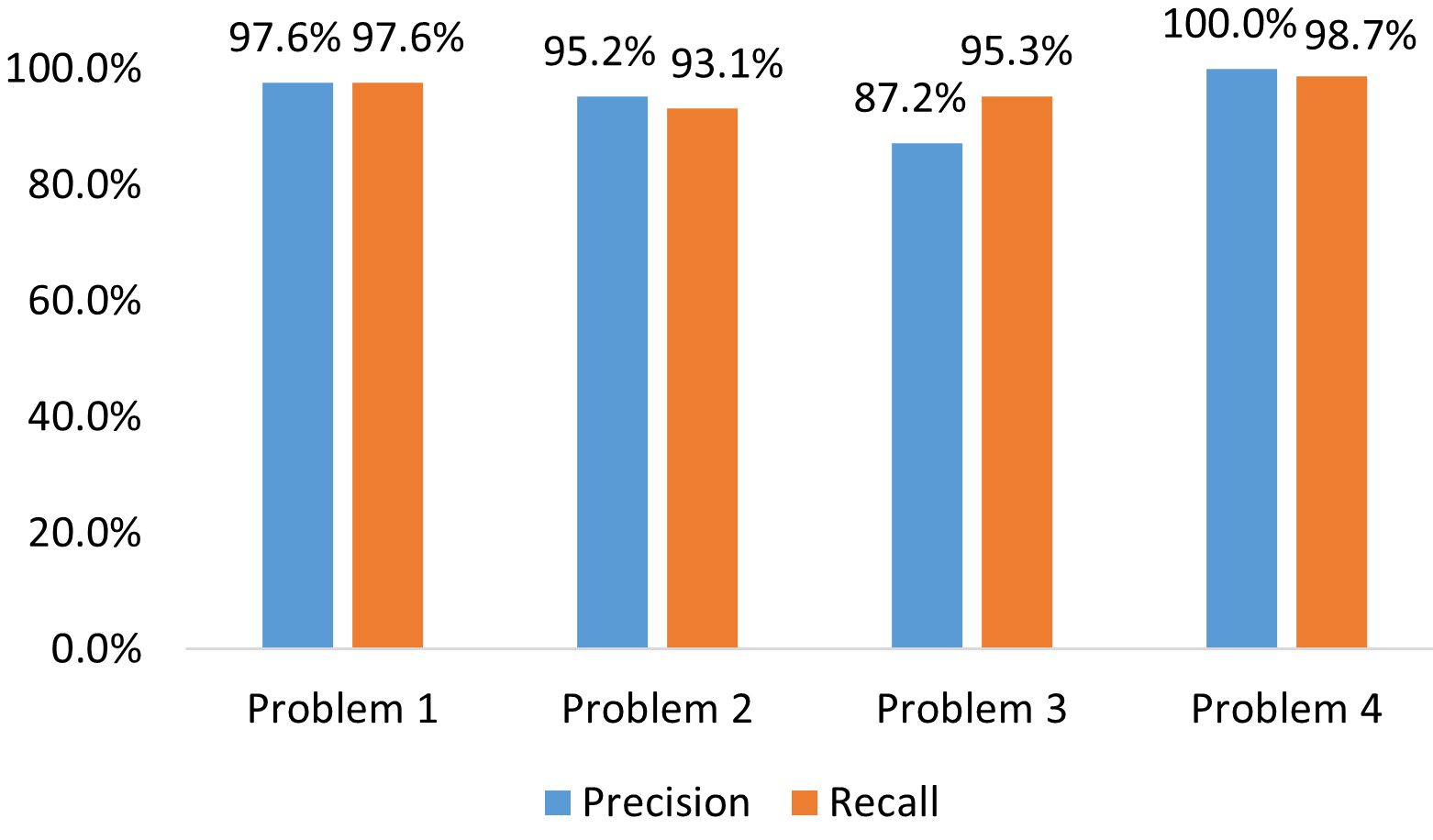}
    \caption{Feedback matching evaluation.}
    \label{fig:table2}
\end{wrapfigure}


To evaluate the effectiveness of our feedback framework in providing accurate and reliable feedback for key instructional points authored by the instructor, we use two metrics: precision (the proportion of selected feedback that was correct) and recall (the proportion of relevant feedback that was successfully selected). For the selected problem, precision and recall were both $97.62$\%, suggesting not only strong alignment when relevant examples were available but also effective coverage of high-priority misconceptions explicitly identified by the instructor. 

To further assess the robustness of our framework, we repeated this protocol on three additional problems from different topics (loops, lists, and strings), each with 100 sampled incorrect submissions. Figure~\ref{fig:table2} reports the precision and recall of matching effectiveness for these problems. These demonstrate that our system shows great potential in effectively delivering accurate and contextually relevant feedback aligned with instructor-defined priorities during active learning sessions, while substantially reducing the risk of propagating incorrect or misleading suggestions to students.

\subsubsection{Explainability of Feedback Propagation.}



\begin{wrapfigure}{r}{0.61\textwidth}
    \vspace{-\baselineskip} 
    \centering
    \includegraphics[width=0.6\textwidth]{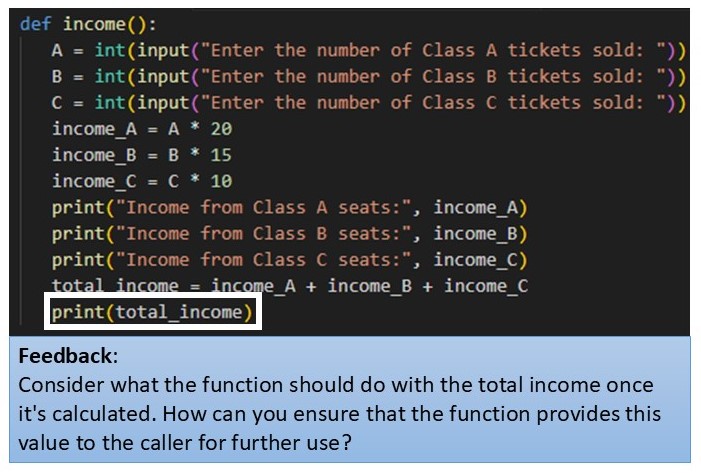}
    \caption{Example of explainable feedback propagation: the highlighted region contributed most to selecting feedback (using \texttt{print} instead of the required \texttt{return}).}
    \label{fig:explainable_feedback}
\end{wrapfigure}


A key advantage of our framework is that each feedback decision is tied to explicit code regions. Because SANN operates over AST subtrees with an attention mechanism, we can highlight the high-attention subtree that most influenced the correctness prediction and the similarity match to an instructor-authored example (Figure~\ref{fig:explainable_feedback}). This makes the feedback pathway inspectable: students see exactly which fragment of code the feedback refers to, and instructors can quickly audit whether the propagated message is appropriate. Although this feature was not integrated into our system during the classroom deployment, it suggests potential for supporting debugging, trust, and iterative refinement of instructor-defined misconceptions. In the future, we plan to integrate this form of localized explanation in both the student interface and the instructor dashboard.


\section{Classroom Deployment}
To complement the offline evaluation, we conducted a preliminary classroom deployment of \textsc{Insight} in an introductory Python programming course at a large university in the Southeastern United States. The system was integrated into an in-class exercise on input/output, variables, arithmetic operations, and functions during week 3 of the course. The instructor used the authoring tool to create five incorrect examples and associated feedback for the exercise. Students completed the activity through the student interface, which provided real-time feedback from the feedback propagation system. The instructor noted that the system was easy to use and that the feedback was clear and accessible for students.

\subsection{Student Perceptions}
A total of 80 students were enrolled in the course, and 69 completed an anonymous post-activity survey about their experience with the system as part of an Institutional Review Board (IRB) approved study protocol. The survey included five statements 
rated on a five-point Likert scale (1 = \emph{Strongly disagree}, 5 = \emph{Strongly agree}): (1) the system provided useful suggestions that helped them revise their work; (2) the feedback was clear and easy to understand; (3) the feedback was specific enough to help them make improvements; (4) the feedback helped them improve their understanding of the course material; and (5) they would like to continue using the system in future coding exercises.

\begin{figure}
    \centering
    \includegraphics[width=\textwidth]{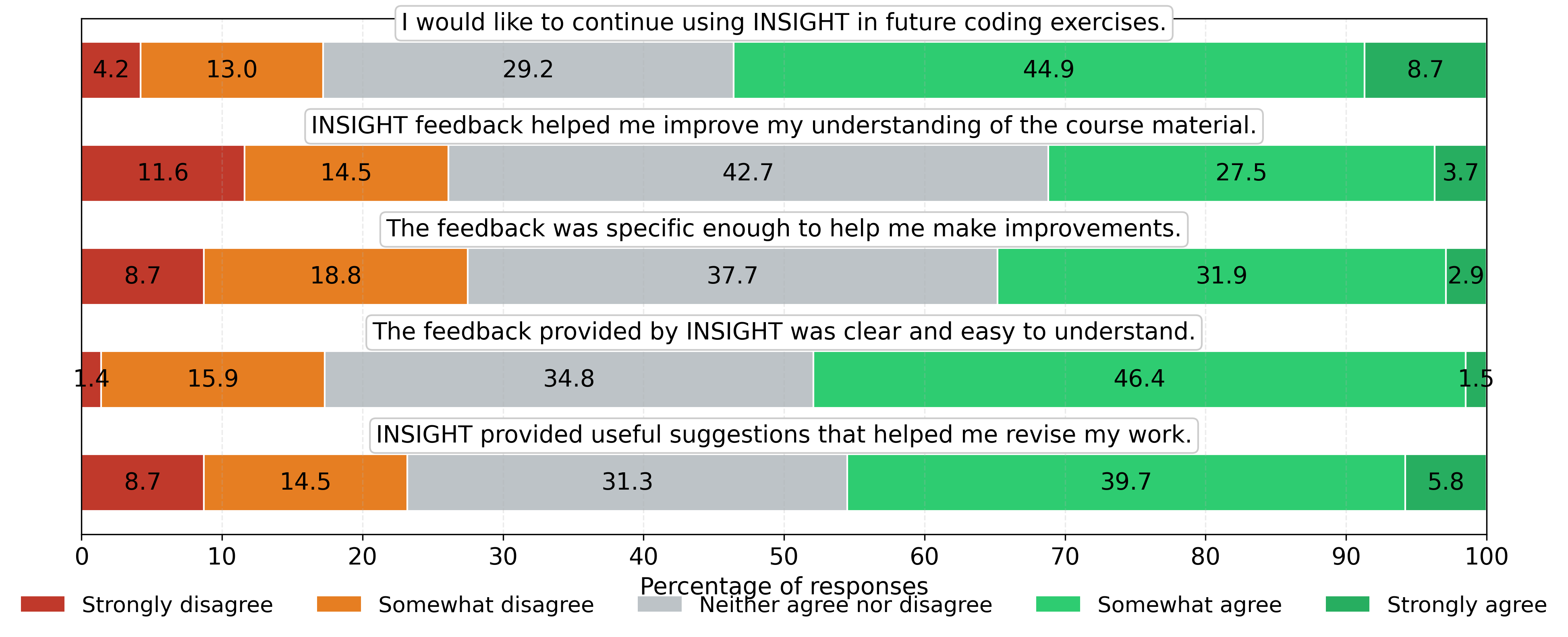}
    \caption{Student perceptions on \textsc{Insight}.}
    \label{fig:survey}
\end{figure}

Students’ ratings ranged from neutral to somewhat agree on all items ($\mu \in [2.98, 3.41]$ on a 5-point scale). For perceived utility, 45.5\% of students somewhat or strongly agreed that the system provided useful suggestions for revising their work, while 23.2\% disagreed, and the remainder were neutral. Perceived clarity was slightly higher: 47.9\% agreed that the feedback was clear and easy to understand, 17.3\% disagreed, and 34.8\% were neutral. Regarding specificity and conceptual support, 34.8\% agreed that the feedback was specific enough to help them improve, and 31.2\% agreed that it helped them better understand the course material, with roughly a quarter of students disagreeing on each item. More than half of the students (53.6\%) reported that they would like to continue using the system in future exercises, whereas 17.2\% disagreed. These results indicate that many students viewed the feedback as helpful and were willing to continue using the system, while a smaller portion found it insufficiently specific or impactful for their understanding. It is also important to consider the system’s current scope: \textsc{Insight} provides feedback only for instructor-identified error patterns, while other issues, such as syntax or runtime errors, are handled by the programming environment or referred to the instructor. Within these constraints, the findings offer initial evidence that the framework is usable and generally well-received, while also pointing to opportunities to improve feedback specificity and expand coverage to better support a wider range of learners.

\subsection{Comparison with GPT-generated Feedback}


\begin{figure}[t]
    \centering
    \includegraphics[width=0.8\columnwidth]{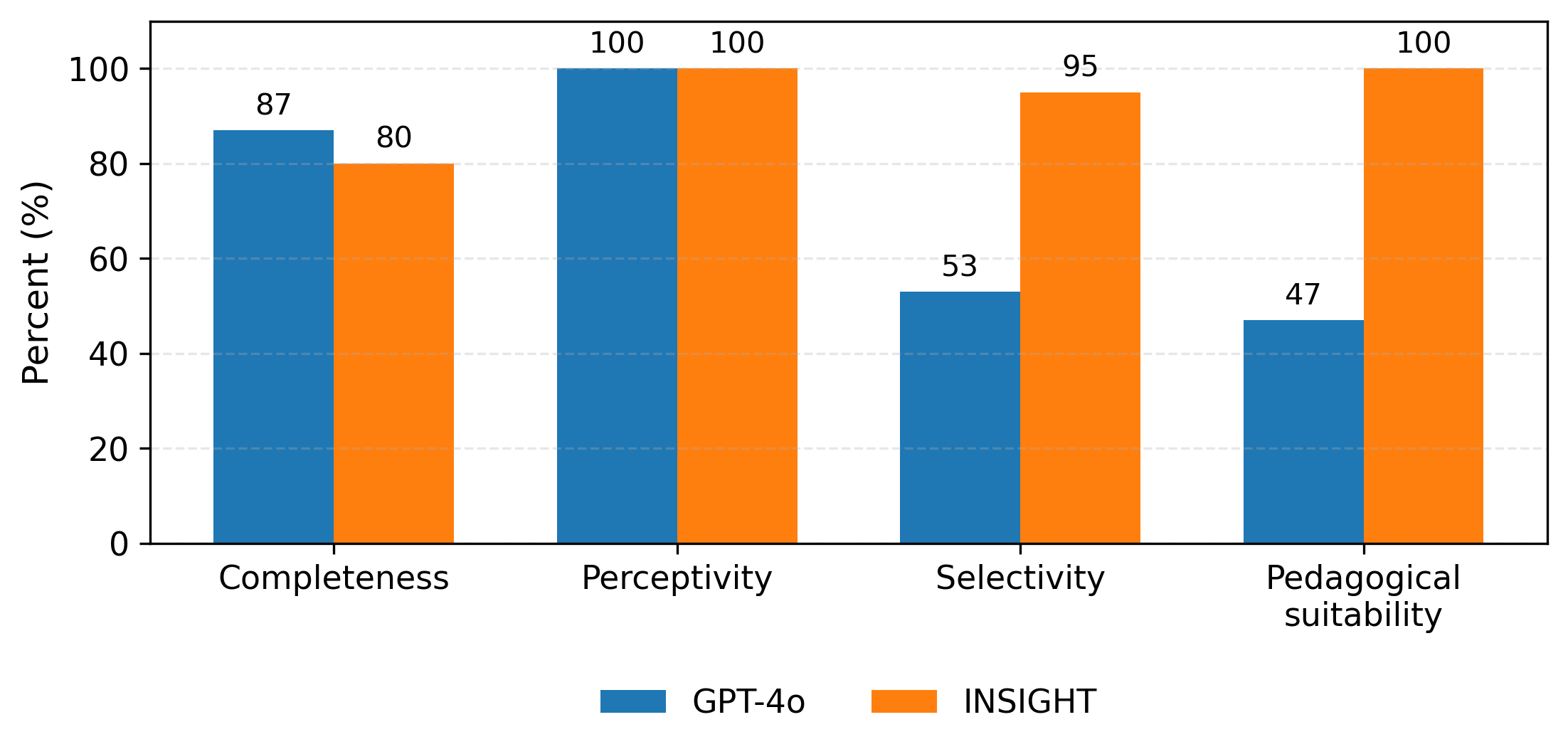}
    \caption{Comparison between feedback from GPT and \textsc{Insight}.}
    \label{fig:gpt_eval}
\end{figure}

To contextualize these findings, we compared feedback from \textsc{Insight} to feedback generated by GPT-4o for all incorrect submissions from the week 3 exercise, focusing on the instructor-defined errors. For each submission, we prompted GPT-4o to produce feedback following the protocol of~\cite{koutcheme2024open}, and then had two human raters ($\kappa$ = $0.84$) evaluate both feedback sources following a codebook~\cite{koutcheme2024open} for completeness, perceptivity, selectivity, and pedagogical suitability, where completeness captures whether all instructor-highlighted issues are addressed, perceptivity whether at least one such issue is identified, selectivity whether no non-existent issues are introduced, and pedagogical suitability whether the feedback offers appropriate guidance without giving away excess information. 

Figure~\ref{fig:gpt_eval} summarizes the human evaluation across completeness, perceptivity, selectivity, and pedagogical suitability. Results show that GPT-4o achieved higher completeness ($87$\% vs. $80$\%), and both approaches reached 100\% perceptivity, indicating that they almost always identified at least one instructor-defined error. However, \textsc{Insight} exhibited higher selectivity ($95$\% vs. $53$\%) and substantially better pedagogical suitability ($100$\% vs. $47$\%), suggesting that GPT-4o was far more prone to introducing extra or misleading issues and to providing overly detailed guidance. In contrast, \textsc{Insight} offered precise, instructor-verified feedback, aligned with instructor-defined error patterns. Together with the classroom survey, these results indicate that our instructor-verified, explainable pipeline provides comparable coverage of relevant issues while more reliably conforming to pedagogical norms for formative feedback.

\section{Discussion}
This work addresses a critical challenge in computer science education: delivering scalable, trustworthy, and pedagogically sound feedback to students in real-time classroom settings to facilitate impactful active learning. As class sizes grow and the need for individualized support increases, traditional feedback mechanisms become difficult to scale~\cite{tang2024sphere}. While LLMs have shown promise in code analysis and feedback generation, concerns around hallucination and reliability hinder their direct deployment in educational environments~\cite{phung2023generative,jacobs2024evaluating,jia2024assessing}. We propose a hybrid framework that integrates an explainable AI model (SANN), instructor-authored feedback, and layered LLM-based verification to deliver technically accurate and pedagogically aligned support. Our feedback propagation approach provides a scalable and trustworthy mechanism for extending instructor-verified feedback to new student solutions, focusing on key instructional concepts and common misconceptions during active learning.

Evaluation results show promising generalization to unseen programming problems, even when the pretrained SANN model is fine-tuned solely on synthetic solutions. The feedback selection pipeline achieves high alignment with instructor intent, despite the inherent challenge of mapping diverse student logic to a limited set of instructor-authored examples. A layered verification mechanism using GPT-4o further reduces incorrect feedback to under 2\%. Together, these components ensure that students receive reliable guidance when their errors align with core instructional points, while less common or previously unidentified issues are addressed by encouraging help-seeking behaviors. Instructors may additionally employ an LLM augmented with instructional guardrails to provide feedback on more open-ended cases beyond predefined misconceptions. This design allows instructors to balance reliability and coverage according to their preferences, while guaranteeing that critical learning issues are consistently addressed with high confidence. Our preliminary classroom deployment further suggests that an explainable, instructor-verified pipeline can be integrated into existing classroom workflows and is generally well-received by students, while still leaving room to improve the specificity and perceived impact of feedback.

Overall, the \textsc{Insight} classroom assistant facilitates effective active learning design by enabling instructor-LLM collaboration in designing active learning activities. A major affordance of this system is offering instructors a tool to scaffold common misconceptions through reliable, scalable, and individualized feedback. The triangulated feedback mechanism ensures that instructor-identified main learning points are reliably conveyed to students, making it well-suited for real-time learning environments. This framework has the potential to empower instructors by reducing the burden of individually assessing each student’s work~\cite{akram2019assessing} and enhancing transparency in automated decision-making, a key requirement in educational AI systems and learning analytics~\cite{hoq2023analysis,hoq2024explaining}.

\section{Limitations and Future Work}

The work has several limitations and important directions for future research. First, the effectiveness of the feedback-matching mechanism depends strongly on the quality and coverage of instructor-authored incorrect solutions, creating a trade-off between instructor authoring time and comprehensive coverage of logical errors in students' code. Our initial interviews have revealed instructors' enthusiasm regarding engaging with \textsc{Insight} to reinforce their key instructional points. A promising direction for future work is therefore to incorporate LLM-based feedback generation specifically for uncovered errors. Following this approach, instructors can spend less time ensuring the most critical misconceptions are reliably covered through instructor-verified feedback, while other errors in students' code are covered by an LLM. Instructors verify unseen error patterns and LLM-generated feedback ad hoc and add them to the repository of verified feedback, thereby expanding coverage while preserving instructional control.


 

Second, our current framework sometimes struggles to identify all issues present in a submission (completeness). Our preliminary analysis showed that once the identified issues are addressed, the system can successfully recall the remaining issues. Future work should include a formal evaluation of the system's ability to retrieve all instructor-identified errors sequentially. It is important to note that sequential, step-by-step, and immediate feedback could support cognitive load management~\cite{anderson1989skill}. Thus, future work will also include studying whether iteratively addressing errors, one at a time, can improve learning outcomes compared to addressing all issues simultaneously. 


Finally, during classroom deployment, while the system matched identified logical errors to instructor-verified feedback, code with runtime or uncovered errors received a generic message prompting students to ask the teaching staff for help. We hypothesize that these “no-feedback’’ cases negatively affected students' reported perceptions of the feedback usability and specificity. Given this limitation, the overall positive perception of students on the AI engine's feedback shows much promise. Furthermore, our current classroom evidence is limited to a single exercise. Future work will involve larger-scale empirical classroom studies to evaluate learning gains, behavioral outcomes, and longer-term usability associated with receiving AI-assisted, instructor-verified feedback. This will ensure that the system is not only technically reliable but also pedagogically impactful.

\section{Conclusion}
This paper has presented a novel student programming feedback framework that combines explainable AI models, LLMs, and instructor expertise to deliver scalable, explainable, and reliable feedback on student programming submissions. By fine-tuning a pretrained model on synthetic data for unseen problems, incorporating LLM-based verification, and centering instructor-identified misconceptions as focal instructional points, the framework supports trustworthy feedback during active learning sessions even in the absence of historical student data. Our evaluation demonstrated the effectiveness of this approach in predicting program correctness and propagating appropriate feedback and an overall positive perception from students in a preliminary classroom implementation. This work lays a foundation for future AI-assisted feedback systems that balance scalability with instructional control and transparency.

\begin{credits}
\subsubsection{\ackname} This research was supported by the National Science Foundation (NSF) under Grants DUE-2236195 and DUE-2331965. Any opinions, findings, and conclusions expressed in this material are those of the authors and do not necessarily reflect the views of the NSF.
\end{credits}
%
%
%
%

\bibliographystyle{splncs04}
\bibliography{references}

\end{document}